\documentclass[aps,showpacs,prc,floatfix,twocolumn]{revtex4}

\usepackage{graphicx}
\usepackage{amssymb}
\usepackage{latexsym}

\newcommand{\cfo}{\mathrm{CFO}}
\newcommand{\fo}{\mathrm{ FO}}
\newcommand{\surf}{d\sigma_\nu}
\newcommand{\beq}{\begin{equation}}
\newcommand{\eeq}[1]{\label{#1} \end{equation}}

\begin{document}

%===================================
\title{
Statistical Hadronization of Supercooled
Quark-Gluon Plasma}
%========================================

\author{A. Ker\"anen}
\email{antti.keranen@oulu.fi}
\affiliation{Department of Physical Sciences,\\
P.O. Box 3000, FIN-90014 University of Oulu, Finland}
\author{L. P. Csernai}
\email{csernai@fi.uib.no}
\affiliation{Section for Theoretical Physics, Department of Physics,\\
University of Bergen, All\'egt. 55, 5007 Bergen, Norway}
\affiliation{KFKI Research Institute for Particle and Nuclear Physics,\\
P.O.Box 49, 1525 Budapest, Hungary}
\author{V. Magas}
\email{vladimir@gtae3.ist.utl.pt}
\affiliation{Center for Physics of Fundamental Interactions
(CFIF), Physics Department, \\
Instituto Superior Tecnico, Av. Rovisco Pais, 
1049-001 Lisbon, Portugal}
\author{J. Manninen}
\email{jaakkoma@paju.oulu.fi}
\affiliation{Department of Physical Sciences,\\
P.O. Box 3000, FIN-90014 University of Oulu, Finland}

\pacs{24.10.Pa, 24.10.Nz, 25.75.Dw, 25.75.Ld}

%\date{}

%%%%%%%%%%%%

\begin{abstract}
The fast simultaneous hadronization and chemical freeze out
of supercooled quark-gluon plasma, created in relativistic
heavy ion collisions, leads to the re-heating of the expanding matter
and to the 
change in a collective flow profile.
We use the assumption of statistical nature of the hadronization
process, and study quantitatively the freeze out in the framework of
hydrodynamical Bjorken model with different quark-gluon plasma
equations of state.
\end{abstract}

\maketitle

%%%%%%%%%%%%%%
\section{Introduction}\label{s:intro}

The hydrodynamical models have been used extensively to study the
evolution of the hot, strongly interacting matter created
in relativistic heavy ion collisions.
These models apply on the space-time region, where the initial, hard
processes have reached the stage where the local 
thermal equilibrium can be assumed, and the strong interactions
between constituent particles are very  frequent.

On the other hand, the thermal statistical models applied
to describe the final hadron abundances have been very successful
for nucleus-nucleus ($AA$)
\cite{features, cleymans319, heppe, cleymans5284, cleymans3319},
 proton-nucleus ($pA$) \cite{cleymans2747} and even for elementary  
$pp$, $p\bar{p}$ and $e^+e^-$ \cite{becattini485,becattini269} reactions.
 The latter ones, particularly, 
suggest the statistical nature of the hadronization process.

In this work, we study the fast hadronization
and chemical freeze-out (CFO)  of locally thermalized
quark gluon plasma (QGP) described by hydrodynamical evolution.
This process is idealized by sudden hadronization over
a three dimensional hypersurface.
The matter crossing the hypersurface is controlled by conservation
of energy momentum- and relevant current conservation laws,
and additionally, by assumption of apparent thermal and
chemical equilibrium 
distribution of the resulting hadron gas (HG).
By {\em apparent} equilibrium distribution we mean that
although there is clearly no room for kinetic equilibration
in elementary collisions -- or in fast, simultaneous phase transition
and CFO in nuclear collisions -- the final hadron spectra are dictated
by maximum entropy bound by conservation of energy
and charge densities.

It is shown, that the first order quasistatic phase transition
is far too slow \cite{CsK92,CsK9} 
to give the hadron chemical decoupling time of
$t_{\cfo} \lesssim 10\ \mathrm{fm}/c$  determined experimentally, see
 \cite{csorgo} and references therein.
Thus, in order to avoid the entropy decrease in the 
non-equilibrium hadronization, QGP must be allowed to supercool,
i.e. to develop mechanical instability before the phase transition
\cite{csorgo}. Although the simultaneous  phase transition and CFO are assumed
to be a non-equilibrium process, we can still exploit
the apparent equilibrium parametrization -- the consequence
of statistical nature of hadronization.

We will show that the shock-like hadronization of supercooled QGP 
leads to the change in collective flow profile, and re-heating
of the system. We study the system
with various choices of the equation of state (EoS)
on both sides of the FO hypersurface.

%%%%%%%%%%%%%%

\section{The Chemical Freeze Out Process}\label{s:fop}

The fast hadronization process, idealized to take place on the
zero-volume hypersurface, leads to a discontinuity in
energy-momentum- and charge conservation equations.
In general, this leads not only to change in density quantities
in LRF, but to the change in flow velocity profile as well
\cite{FO1,FO3}.

Denoting post freeze out quantities by subscript $\fo$, 
the energy-momentum ($T$) conservation at the surface element
$\surf$ leads to a set of four equations 
\begin{equation}\label{eq:Tcons}
T^{\mu\nu}\surf = T_\fo^{\mu\nu}\surf.
\end{equation}
In the isospin symmetric case, the electric charge conservation
is trivial, and we are left with local baryon number and strangeness
conservation equations:
\begin{eqnarray}
N^{\nu}_B\surf &=& N^{\nu}_{B,\fo}\surf \label{eq:Bcons}  \\
N^{\nu}_S\surf &=& N^{\nu}_{S,\fo}\surf \label{eq:Scons}.
\end{eqnarray}

Here, we restrict our considerations to time-like FO hypersurfaces
in order to avoid currents entering from post-FO to pre-FO side.
The complications in the general case, where the space-like
FO is included, are discussed extensively in \cite{FO1,FO3}. 
For the time-like hypersurface, there is always a proper
Lorentz transformation for each point $x$ such that
solving the equations (\ref{eq:Tcons},\ref{eq:Bcons},\ref{eq:Scons}) can be
carried out in a frame, where flow velocities are
$u(x) = \gamma(1,0,0,v)$ and $u_\fo(x) = \gamma_\fo(1,0,0,v_\fo)$
on pre- and post-FO sides, respectively. After solving the FO equations,
the actual flow velocity $u_\fo=\gamma_\fo(1,\vec{v}_\fo)$
 is obtained by a simple spatial rotation.
Provided with these tools, only two of the four equations
(\ref{eq:Tcons}) are independent.

Usually, the strangeness in hydrodynamical simulations is assumed to
be homogeneously distributed, so the condition $n_S(x) = n_{S,\fo}(x)=0$
binds the parameters associated with strangeness conservation.

The process described is generally non-adiabatic, but the
entropy constraint
\begin{equation}\label{eq:entr}
S^\nu\surf\leq S^\nu_\fo\surf
\end{equation}
must be satisfied in every point on the FO hypersurface.

%%%%%%%%%%%%
\subsection{The Statistical Post Freeze Out Hadronization}

Statistical models were applied
for high energy collisions \cite{Fe:50, Po:51, La:53}
from the beginning of this field.
In the last decade a significant development of these
models and the extension of the area of their
applicability took place.
The main reason for this is a surprising success
of the statistical approach in reproducing new experimental data
on hadron multiplicities in
nuclear ($AA$) \cite{features, cleymans319,heppe, cleymans5284, cleymans3319}
 and elementary ($e^+e^-$, $pp$,
$p\bar{p}$) collisions \cite{becattini485,becattini269}.
One of the important results of the analysis of hadron yield
systematics at high energies (SPS and higher)
done within the statistical models
is the approximate independence
of the temperature parameter ($T = 160\div 180$~MeV)
of the system size and collision energy.
Another important feature found in statistical analyses is the
common freeze-out condition, energy per hadron
$E/N \sim 1$ GeV, found for all systems from GSI 2 GeV A NiNi
 to SPS 160 GeV A PbPb collisions \cite{jeanPRL,features}.
These results can be attributed to
the statistical character of the hadronization process.

Calculations within statistical models are straightforward
when the mean number of particles of interest is large
and consequently it is enough to fulfill the conservation
laws in the average sense, i.e. the grand canonical (GC) 
description can be used. In GC ensemble,
the mean particle multiplicities
are just proportional to the volume $V$ of the system.
Thus, the particle densities and the ratios
of the multiplicities of two
different particle species, which are usually
used for the comparison with the data,
are volume independent.

This simple volume dependence is however not valid
any more for a small system
in which the  mean particle multiplicity is low, like
in $pp$ collisions.
In this case
the material conservation laws
should be imposed exactly on each charge configuration
 of the system, i.e. canonical ensemble (CE) description should be used.
This condition introduces a significant correlation
between particles
who carry conserved charges (see \cite{keran,kerabeca},
and references therein).

In PbPb collisions at the SPS collider and AuAu reactions at RHIC we
produce large and hot systems,
where the corrections are negligible and GC
description is satisfactory for most of the produced particles.
Nevertheless,
the number of strange particles created is rather low, and the net strangeness
is exactly zero.
Therefore the CE 
description would be preferable for the strange particles (we
are not going to include particles with charm or more heavy quarks).
However, the basic assumption of the hydrodynamics is the
{\em local} thermal and chemical equilibrium, whereas the canonical
 conservation
laws can only be realized {\em globally}. This prevents us using
CE in  post FO, which
is nothing but the extension of the hydrodynamical part with
different EoS.
This drawback is not of a serious concern, because the conditions
reached in high energy heavy ion systems are expected to fulfill 
the requirements for GC description \cite{kerabeca}.

In the relativistic GC ensemble for ideal hadron gas (HG), the properties
of the matter in unit volume are 
parametrized by temperature $T$ and
fugacities $\lambda_i$ (or chemical potentials $\mu_i$)
for conserved charges.
For example, the pressure and energy density are
 $P=P(T,\lambda_B,\lambda_S)$ and  $e=e(T,\lambda_B,\lambda_S)$
if we omit
the fugacity $\lambda_Q$ associated with electric charge conservation.
%We calculate the energy-, charge- and entropy densities as
%sums over partial densities due to listed hadrons \cite{pdg}
%up to the mass of 2 GeV.  

In case of sudden freeze out there might not be sufficient time to
achieve chemical equilibration of number of strange- and anti-strange
quarks. Thus, an over- or underpopulation of strange hadron species
may persist after hadronization. This can, however, be treated in the
GC approach as we will see.

Now that the thermal ideal gas of hadrons provides us with 
tools to calculate post-FO quantities, we can collect the
set of equations needed to describe the simultaneous FO and hadronization
taking place at a time-like hypersurface.
For the sake of simplicity, we choose a hypersurface of constant
coordinate time, $\surf = (1,0,0,0)$, which is, of course, 
connected with any time-like hypersurface by a proper Lorentz transformation.
Recalling the energy-momentum, baryon current and strangeness
conservation equations yields
\begin{eqnarray}
(e+P)\gamma^2 - P &=& (e_\fo +P_\fo)\gamma_\fo^2 - P_\fo 
\label{T00} \\
(e+P)v\gamma^2 &=& (e_\fo +P_\fo)v_\fo\gamma_\fo^2
\label{T30} \\
n_B\gamma &=& n_{B,\fo}\gamma_\fo
\label{nbc} \\
n_S\gamma &=& n_{S,\fo}\gamma_\fo = 0 \label{nsc}.
\end{eqnarray}
Given the pre-FO quantities, the solution of this set of equations
gives the post-FO parameters:
$v_\fo,\ T_\fo,\ \lambda_{B,\fo}$ and  $\lambda_{S,\fo}$,
flow velocity of the fluid element, temperature and fugacities
for conserved currents, respectively.
These parameters describe completely the hadron spectra and other
statistical quantities at the LRF of the fluid element.

In order to take into account the possible over- or underpopulation of
strange- and anti-strange quarks in post-FO side, 
a parametrization of the conservation of the number of
$s,\bar{s}$ pairs must be introduced.
In the GC formulation, a conservation law gives rise to a fugacity parameter.
The one associated with the number of $s,\bar{s}$ pairs,
$n_{s,\bar{s}}$, is usually
called $\gamma_S$, first introduced in Ref. \cite{gammas}.
The assumption of the survival of $n_{s,\bar{s}}$ over the
FO hypersurface yields yet another equation to be included in the set above:
\begin{equation} \label{nss}
n_{s,\bar{s}}\gamma = \left(n_{s,\bar{s}}\right)_\fo\gamma_\fo.
\end{equation}
 
%%%%%%%%%%%%%%%%%%

\section{Numerical Studies}

For quantitative studies of the FO, we have chosen a framework of
Bjorken model \cite{bjorken} for the expanding QGP.
This allows us to cover many FO scenarios with relative ease.
Within the Bjorken model, the evolution of matter in
one spatial dimension $z$ and charge densities $n_i$
is governed by equations
\begin{eqnarray}
        \frac{\partial e}{\partial \tau} + \frac{e+P}{\tau}&=&0
\label{bmotion} \\
\label{bmotion2}
        \frac{\partial P}{\partial y} &=&0 \\
\frac{\partial n_i}{\partial \tau} + \frac{n_i}{\tau}&=&0,
\label{bmotion3}
\end{eqnarray}
where $\tau=\sqrt{t^2-z^2}$ is the proper time and
$y$ is the rapidity of a given fluid element.
For vanishing net charges, thermodynamical quantities
are constant along constant proper time curves on $(t,z)$ plane.
Taking the FO to take place at constant {\em coordinate} time,
the proper time of freezing out fluid element decreases with
increasing spatial coordinate, so the temperature is an
increasing function of $z$.
The constant coordinate time choice is made in order to have a clear picture
of the FO process in different thermal circumstances. 
In addition to the equations above, one needs an equation
to describe the thermodynamical structure of the evolving
matter, the equation of state.

%%%%%%%%%%%%%%
\subsection{The Equation of State - Supercooled QGP}

Three different equations of state for supercooled 
quark-gluon plasma are considered.
The QGP is assumed to be an ideal gas of
three flavors (u,d,s) of quarks and their antiquarks.
We will model the thermodynamics of the QCD vacuum in terms of the MIT
bag model ~\cite{jaffe:origBagmodel} by introducing a phenomenological
bag constant $B$. In this model, confinement can be thought of as a rule which
allows quarks to inhabit only regions of a ''false vacuum'', which has
minimum energy density $B$. This is due to 
the fact that quarks can not be found
in physical vacuum but only inside hadrons that are small bags of
the false vacuum containing triplets or pairs of quarks. In QGP, all 
the space is
endowed with this energy density $B$. 
%Estimates of the magnitude of bag constant vary,
%but it is  usually restricted to be in range $B^{\frac{1}{4}}=[0,0.3]$ GeV.

Light quarks ($u$,$\bar{u}$,$d$,$\bar{d}$) can be considered as massless,
 but the mass of strange quark can not
be neglected. For massless quarks and gluons, integral in partition function
$Z_{\mathrm{QGP}}$ can be evaluated
analytically, but the contribution of strangeness must be calculated 
numerically.
It is convenient to split $Z_{\mathrm{QGP}}$ into four parts:
\begin{displaymath}
        \ln Z_{\mathrm{QGP}} = \ln Z_f + \ln Z_b + \ln Z_s + \ln Z_{vac},
\end{displaymath}
where $ \ln Z_f  $, $ \ln Z_b $ and $ \ln Z_s $ are the contributions 
of massless quarks,
gluons and massive strange quarks respectively.
$ \ln Z_{vac} $ is the contribution of the bag constant added to the 
energy-momentum
tensor. Evaluation of analytical parts of $\ln Z_{\mathrm{QGP}}$ leads 
to the expressions
~\cite{suhonen:QGP}
\begin{eqnarray}
        \ln Z_f &=& V (\frac{7}{30}\pi^2 T^3
        + \mu_q^2 T + \frac{1}{2 \pi^2} \mu_q^4\frac{1}{T}) \\
        \ln Z_b &=& \frac{8}{45}\pi² V T^3 \\
        \ln Z_{vac} &=& -\frac{BV}{T},
\end{eqnarray}
where  $ \mu_q $ is a light quark chemical potential,
$ \mu_q = \mu_B/3 $.
The total strangeness is zero during the whole collision evolution.
If one assumes that the strangeness produced in a collision is spread
homogeneously
in QGP, then $ n_S $ must be zero in any given fluid element,
so the contribution
of $s$ and $\bar{s}$ quarks is
\begin{equation}\label{lnZs}
        \ln Z_s=\frac{6V}{\pi^2}\int_{0}^{\infty} dp\,
        p^2  \ln \left( 1 + e^{-\beta \sqrt{p^2+m_s^2}} \right),
\end{equation}
where $m_s$ is the mass of the strange quark, here chosen to
vary between 150 and 250 MeV.

If one puts the bag constant to zero (B$\equiv$0), then $\ln Z_{vac}$ 
term vanishes and we end
up with the normal quantum distribution for the ideal gas. 
In this case, the explicit form
of the three flavor QGP EoS  is\footnote{The massive $s,\bar{s}$ quarks
give a little correction to this simple rule, but we suppress that 
in the text.}
\beq
        P(T,\mu_i) = \frac{1}{3}e, \quad  e(T,\mu_i) = e_{SB}.
\eeq{IdEoS}
In the case of finite bag constant, EoS reads:
\beq
        P(T,\mu_B,B) = \frac{1}{3}e_{SB} -B, \quad  e(T,\mu_B) = e_{SB} + B.
\eeq{BagEoS}

The third equation of state in addition to ones
with finite and zero MIT bag constants considered here was suggested
in the Ref.
~\cite{jenk,jenk2}
(and first time applied for heavy ion collisions in ~\cite{jenkjenk})
for the case of
baryon free QGP ($\mu_q=0$) with two quark flavors:
\beq
P(T)=\frac{1}{3}e(T) - bT, \quad e(T)=e_{SB}.
\eeq{jeos}
We call this a {\em spinodal EoS}, inspired by the existence of a local
minimum in pressure profile (\ref{jeos}).
It is easy to check that $e$ and $P$ are connected by the
thermodynamical equation
\beq
e=T\frac{dP}{dT}-P.
\eeq{bT}
This EoS can be justified as resulting from the non-perturbative QCD effects:
The lattice calculations (see for example ~\cite{blum:QCD}) show that close
to the critical temperature for hadronic matter, QGP phase transition
energy density has a sharp increase and soon saturates with an
equilibrium Stefan-Boltzmann value $e(T)=e_{SB}(T)$, while pressure
increases much more slowly and reaches Stefan-Boltzmann limit only at
very high $T$.
Such a pressure suppression is presented by eqs. (\ref{jeos}).

The spinodal EoS considered here will be of form (\ref{jeos}) but with
three quark flavors
and finite net baryon density:
\beq
P(T,\mu_i)=\frac{1}{3}e(T,\mu_i) - bT, \quad e(T,\mu_i)=e_{SB}.
\eeq{btEoS}
Comparing these expressions with the bag model EoS one can
see that for the pressure the bag constant, $B$, is replaced by T and
$\mu$  dependent
term $bT$, while for energy density the analogy
with bag model disappears due to the relation (\ref{bT}).

If the phase transition temperature $T_c$ is known, parameter $b$ can be
fixed from the Gibbs condition $P_{HG}(T_c,\mu)=P_{QGP}(T_c,\mu)$. 
This same condition fixes the constant $B$ within the bag model.

\subsection{Freeze Out Illustrated}

The post-FO matter is described by the quantum ideal gas,
composed of
hadrons up to a mass of $2.5$ GeV, 
listed in the year 2000 issue of Review of Particle Physics
\cite{pdg}.
In order to satisfy the conservation equation (\ref{nss}) for
strange quark pairs, additional fugacity $\gamma_S$ is introduced
for each strange particle $i$ as $\lambda_i \rightarrow 
\lambda_i \gamma_S^{|S_i|}$. 
Additionally, the mesons carrying $s\bar{s}$ pairs must be taken
into account. In this work, the number of meson $i$ is affected
by the fugacity factor $\lambda_i = \gamma_S^{2c_s}$,
where $c_s$ is the relative $s\bar{s}$ content in the meson.
We take $c_s = 0.5$ for $\eta$ mesons, and  $c_s = 1$ for
$\phi$, $f_0(980)$, $f_1(1510)$, $\phi(1680)$ and $\phi_3(1850)$.   

\begin{figure}
        \centering
        \includegraphics[scale=0.7]{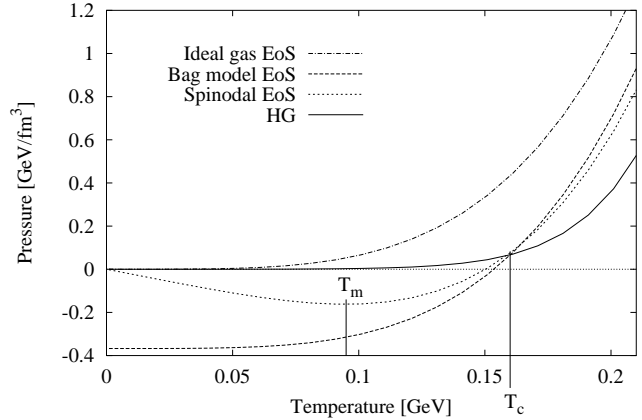}
        \caption{Hadron gas pressure and
	quark-gluon plasma pressures with ideal gas,
        bag model and spinodal model equations of state with parameters
        $\mu_B=100$ MeV and $T_c=160$ MeV.}
        \label{fig:criticalT}
\end{figure}
In figure \ref{fig:criticalT} we show the pressures of the HG and QGP
as functions of temperature. The parameters appearing in different
equations of state for QGP are fixed by setting the critical temperature
to $T_c = 160$ MeV. The pressure minimum in the spinodal EoS is labeled
by $T_m$. 
Of course, the critical temperature varies with baryon density
(or $\mu_B$), but we find this variation to be negligible
within reasonable range of FO density.  
The strangeness saturation parameter $\gamma_S$ is set to
one, corresponding
to the full strangeness equilibration.
We always let the QGP side be in full strangeness equilibrium,
so the Gibbs condition for the adiabatic, isothermic phase transition 
compels $\gamma_S = 1$.
In the following, however, the phase transition is neither isothermic
nor adiabatic,
so the change in fugacities is allowed and unavoidable.

\begin{figure}
        \centering
        \includegraphics[scale=0.7, angle=0]{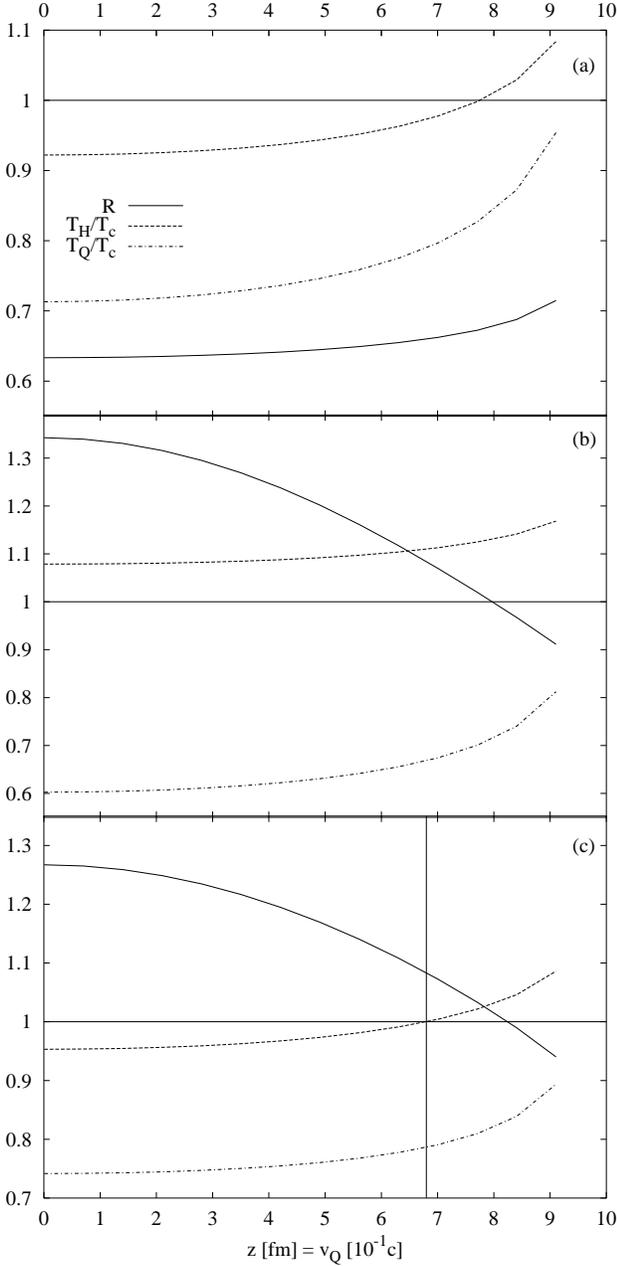}
	\caption{Various quantities on the FO surface.
	Panels from top to bottom: a) ideal gas EoS for
	QGP, b) MIT bag EoS, c) spinodal EoS.
	R is the ratio of entropy currents in
	HG and QGP, $T_H$ and $T_Q$ are the temperatures
	on the HG and QGP side, and $T_c$ is the critical temperature.
	The model parameters
	in the bag- and spinodal EoS are fixed to produce
	critical temperature $160$ MeV. For the ideal gas EoS,
	there is no real $T_c$, the chosen $160$ MeV is arbitrary.
	The vertical line in the panel c marks the 
	point, where the HG temperature crosses the critical value.}
	\label{fig:2}
\end{figure}
\begin{figure}
        \centering
        \includegraphics[scale=0.7, angle=0]{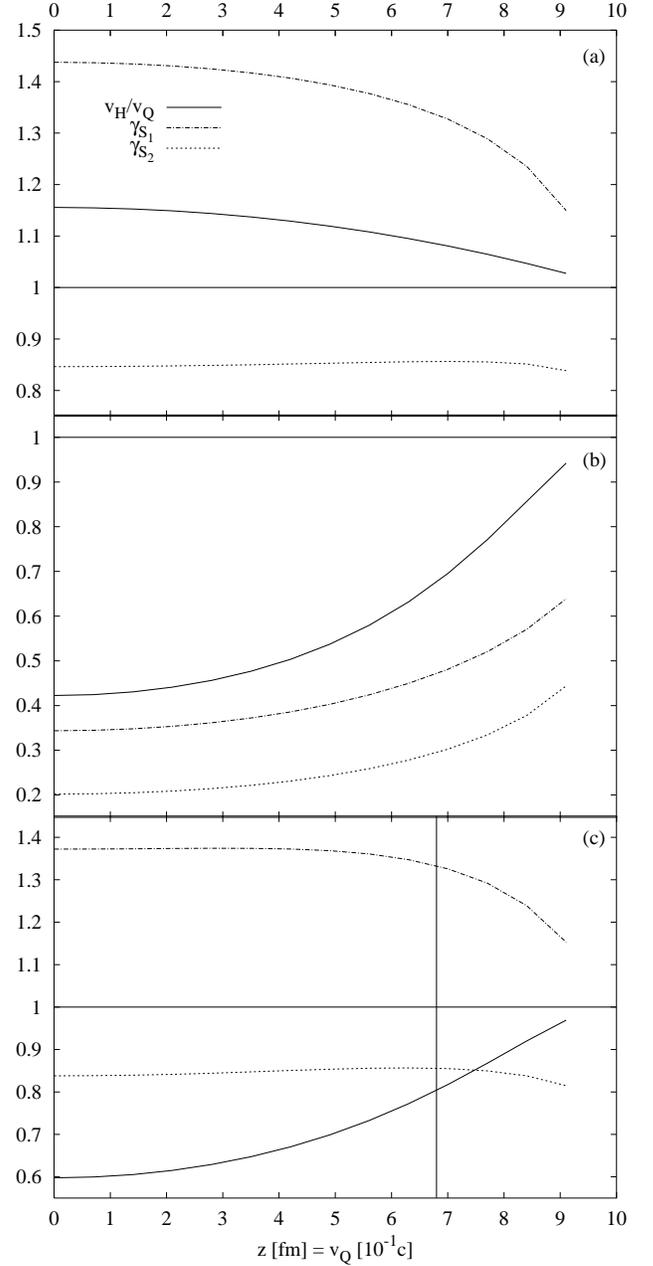}
	\caption{Various quantities on the FO surface.
	Panels from top to bottom: a) ideal gas EoS for
	QGP, b) MIT bag EoS, c) spinodal EoS.
	$v_H$ and $v_Q$ are the HG and QGP flow velocities.
	$\gamma_{S_1}$ and $\gamma_{S_2}$ are the strangeness
	saturation fugacities with $s$ quark masses 
	$150$ MeV and $250$ MeV, respectively. The model parameters
	in the bag- and spinodal EoS are fixed to produce
	critical temperature $160$ MeV. For the ideal gas EoS,
	there is no real $T_c$, the chosen $160$ MeV is arbitrary.
	The vertical line in the panel c marks the 
	point, where the HG temperature crosses the critical value.}
	\label{fig:3}
\end{figure}

The figure \ref{fig:2} depicts the variation of some relevant 
variables along the coordinate time $t=10$ fm$/c$ in the Bjorken
model. Going from the edge of the light cone ($z = 10$ fm) to
the non-flowing center of the system ($z = 0$ fm), we describe
the changes of quantities in proper time of the given fluid element from
$\tau = 0$ to $\tau = 10$. This is, going from the origin 
to the right, we go from the later, cooler and more dilute
stage of evolution to the earlier, hotter and more dense stage.
We have chosen the initial values for energy- and
baryon density $e_0 = 5$ GeV fm$^{-3}$ and $(n_B)_0 = 1$ fm$^{-3}$,
respectively, at the time $\tau=1$ fm$/c$. For the ideal gas QGP EoS,
the higher $e_0 = 8$ GeV is used in order to keep the figure
more descriptive.
The model parameters $B$ in the MIT bag EoS and $b$
in the spinodal EoS are fixed to produce 
critical temperature $160$ MeV, while for the ideal gas EoS there
is no critical temperature. The arbitrary value $T_c = 160$ MeV
is chosen to give similar scaling for temperatures in all cases.
The ratio $R$ of entropy currents in HG and QGP stays well above
one for bag and spinodal models all the way up to
unrealistic FO conditions, so the substantial 
increase in entropy is expected in fast simultaneous FO and
hadronization. For the reference, the $R$ for the ideal gas QGP
is, as expected, far below one.
The crossing of HG temperature $T_H$ with $T_c$ marks the endpoint
of the physically allowed FO. For the bag model, there is no such
crossing, but the FO stays unphysical with all reasonable values
of QGP temperature $T_Q$. We could choose lower initial value
for the QGP energy density, but the the degree of supercooling,
$T_Q/T_c$, would be of order $0.5$ at the HG critical point.
For the spinodal model QGP, the case is different.
At the point $T_H/T_c = 1$, the degree of QGP supercooling
is about $24\%$, and stays rather constant when going towards
the cooler system.

In figure \ref{fig:3} we illustrate the changes in flow velocities
and strangeness saturation parameter in the circumstances
equal to ones in figure \ref{fig:2}.
For both cases, bag- and spinodal QGP, the flow is decelerated
at the FO to HG, indicated by the ratio of HG and QGP velocities,
$v_H/v_Q$. At the point $T_H = 160$ MeV for the spinodal EoS,
this ratio is $0.80$, indicating the final HG flow velocity
of 0.54$c$.  
The $\gamma_S$, resulting from the survival of $s\bar{s}$ pairs
through the FO process, is calculated for two different values
of $s$ quark mass $m_s$. $\gamma_{S_1}$ corresponds to
$m_s = 150$ MeV and $\gamma_{S_2}$ is for
$m_s = 250$ MeV. Comparing the figures, we find $\gamma_S$ 
to be very sensitive to the choice of QGP EoS and the $m_s$.
%With the more reasonable value, $m_s = 150$ MeV, $\gamma_S = 0.74$ 
%at the point $T_H/T_c = 1$ for the spinodal EoS. This value is
%similar to one found in statistical model fit to NA49 Pb-Pb data
%\cite{features}.
It is worth noting, that fixing $\gamma_S$ to one or varying the
$m_s$ gives only negligible change to other quantities in
figures \ref{fig:2} and \ref{fig:3}.

%%%%%%%%%%%%%%%%%
\section{Conclusions}

We have found that the realistic and accurate study of the freeze-out
process is important and cannot be neglected.   Our results show that
the FO process is very sensitive on the properties of the
EoS.  Since our main goal is to identify the EoS from the data this
is an observation of basic importance.

The correct treatment shows that FO is not even possible  from arbitrary
kind of initial state, and  the entropy constraint is a sensitive  condition.

From the point of strangeness, we can also conclude that
strangeness production is very sensitive to the correct FO treatment
and to the pre FO EoS. Thus, strangeness data not only provide
a signal of QGP formation, but with proper and realistic description of
freeze out, strangeness provides the most sensitive signal  indicating
different properties of the pre FO EoS.

In conclusion, this study shows that  collective, continuum reaction 
models, like fluid dynamical (FD)
models (one fluid FD, multi - fluid FD, chiral FD), 
can and must be supplemented with realistic Freeze Out treatment
to evaluate measurable data.  These calculations indicate that this
is now possible cell by cell in FD models, although it requires 
more than average computational capacity, and needs preferably high
performance parallel computing.

%%%%%%%%%%%%
\section*{Acknowledgments}

We acknowledge the constructive discussions with Esko Suhonen. 
Two of us (A. K. and V. M.)
acknowledge the support of the Bergen Computational
Physics Laboratory in the framework of the European Community - Access
to Research Infrastructure action of the
Improving Human Potential Programme.

%%%%%%%%%%%%%%

%%%%%%%%%%%%


\begin{thebibliography}{99}
\bibitem{features}
F. Becattini, J. Cleymans, A. Ker\"anen, E. Suhonen and
K. Redlich,
{\em Phys. Rev.} {\bf C64} (2001) 024901

\bibitem{cleymans319}
J. Cleymans, D. Elliott, R.L. Thews and H. Satz,
{\em Z. Phys.} {\bf C74} (1997) 319

\bibitem{heppe} P. Braun-Munzinger, I. Heppe and J. Stachel,
{\em Phys. Lett.} {\bf B465} (1999) 15

\bibitem{cleymans5284}
J. Cleymans and K. Redlich, {\em Phys. Rev. Lett.} {\bf 81} (1998) 5284

\bibitem{cleymans3319} J. Cleymans, D. Elliott, A. Ker\"anen,
E. Suhonen, {\em Phys. Rev.} {\bf C57} (1998) 3319

\bibitem{cleymans2747}
J. Cleymans, A. Ker\"anen, M. Marais and E. Suhonen,
{\em Phys. Rev.}  {\bf C56} (1997) 2747

\bibitem{becattini485}  F. Becattini, {\em Z. Phys.} {\bf C69} (1996) 485

\bibitem{becattini269}
F. Becattini and U.~Heinz, {\em Z. Phys.}  {\bf C76} (1997) 269

\bibitem{CsK92}
L.P. Csernai, J.I. Kapusta,
{\em Phys. Rev.} {\bf D 46} (1992) 1379

\bibitem{CsK9}
L.P. Csernai, J.I. Kapusta,
{\em Phys. Rev. Lett.} {\bf 69} (1992) 737

\bibitem{csorgo}
T. Cs{\"o}rg\H o and L.P. Csernai, {\it Phys. Lett.} {\bf B 333} (1994) 494

\bibitem{FO1}
Cs. Anderlik {\it et al}, {\it Phys. Rev.} {\bf C 59} (1999) 388;
{\it Phys. Rev.} {\bf C 59} (1999) 3309;
{\it Phys. Lett.} {\bf B 459} (1999) 33
\bibitem{FO3}
V.K. Magas {\it et al}, Heavy Ion Phys. {\bf 9} (1999) 193, (nucl-th/9903045);
Nucl. Phys. {\bf A 661} (1999) 596

\bibitem{Fe:50}
E. Fermi, {\it Prog. Theor. Phys.} {\bf 5} (1950) 570

\bibitem{Po:51}
I. Ya. Pomeranchuk, Dokl. Akad. Nauk SSSR
{\bf 78} (1951) 884

\bibitem{La:53}
L. D. Landau, {\it Izv. Akad. Nauk SSSR}, Ser. Fiz.
{\bf 17} (1953) 51

\bibitem{jeanPRL}
J. Cleymans and K. Redlich, {\em Phys. Rev. Lett.} {\bf 81} (1998) 5284

\bibitem{keran}
J. Cleymans, A. Keranen and E. Suhonen,
presented at  the
11th Chris Engelbrecht Summer School in Theoretical Physics,
``Hadrons in Dense Matter and Hadrosynthesis'',
Cape Town 4-13 February 1998, Springer-Verlag Berlin Heidelberg 1999,
hep-ph/9809261

\bibitem{kerabeca}
A. Ker\"anen and F. Becattini,
{\it Phys.~Rev.} {\bf C65} 044901 (2002)

\bibitem{gammas} P. Koch, B. M\"uller and J. Rafelski,
{\em Phys. Rep.} {\bf 142}, (1986) 167

\bibitem{bjorken}
J. D. Bjorken, {\em Phys. Rev.} {\bf D27} (1983) 140

\bibitem{jaffe:origBagmodel}
A. Chodos, R. L. Jaffe, K. Johnson and C. B. Thorn,
{\em Phys. Rev.} {\bf D10} (1974) 2599

\bibitem{suhonen:QGP}
J. Cleymans, R. V. Gavai and E. Suhonen,
{\em Phys. Rep.} {\bf 130} (1986) 217

\bibitem{jenk} C.G. Kallman,
{\it Phys. Lett.}
{\bf B 134} (1984) 363

\bibitem{jenk2} M.I. Gorenstein, O.A. Mogilevsky,
{\it Z. Phys.} {\bf C 38} (1988) 161

\bibitem{jenkjenk} L.L. Jenkovszky, B. Kampfer, V.M. Sysoev, {\it Z. Phys.}
{\bf C 48} (1990) 147

\bibitem{blum:QCD}
T. Blum, L. K\"arkk\"ainen and D.Toussaint,
{\em Phys. Rev.} {\bf D51} (1995) 5153

\bibitem{pdg} 
D. E. Groom {\em et al}, {\em Eur. Phys. J.} {\bf C15} (2000) 1


\end{thebibliography}
\end{document}